\documentstyle[11pt,newpasp,twoside,epsf]{article}
\def\edcomment#1{\iffalse\marginpar{\raggedright\sl#1\/}\else\relax\fi}
\reversemarginpar

\begin{document}
\title{Clouds and Clearings in the Atmospheres of the L and T Dwarfs}
 \author{M.S. Marley}
\affil{Space Sciences Division, Mail Stop 245-3, NASA Ames Research Center,
Moffett Field, CA 94035}
\author{A. Ackerman}
\affil{Earth Sciences Division, Mail Stop 245-4, NASA Ames Research Center,
Moffett Field, CA 94035}
\author{A.J. Burgasser}
\affil{Hubble Fellow, Department of Astronomy \& Astrophysics, UCLA, Los Angeles, CA 90095-1562}
\author{D. Saumon}
\affil{Department of Physics \& Astronomy, Vanderbilt University, Nashville,
TN 37235}
\author{K. Lodders}
\affil{Dept. of Earth \& Planetary Science, Washington University, St. Louis,
MO 63130}
\author{R. Freedman}
\affil{SETI Institute, NASA Ames Research Center, Moffett Field, CA 94035}

\begin{abstract}
A sophisticated approach to condensate opacity is required to properly model
the atmospheres of L and T dwarfs.  Here we review different models
for the treatment of condensates in brown dwarf atmospheres.  We conclude
that models which include both particle sedimentation and upwards
transport of condensate (both gas and particles) provide the best fit
for the L dwarf colors.  While a globally uniform cloud model fits the
L dwarf data, it turns to the blue in $J-K$ too slowly to fit the T dwarfs.
Models which include local clearings in the global cloud deck, similar
to Jupiter's prominent five-micron hot spots, better reproduce the
available photometric data and also account for the observed resurgence
of FeH absorption in early type T dwarfs.

\end{abstract}

\section{Introduction}

Long before the first discoveries of brown dwarfs, it was
recognized that condensates would play a critical
role in controlling their atmospheric opacity,
at least in certain effective temperature ranges
(Stevenson 1986; Lunine et al. 1989).  It was also evident that the correct 
choice for the vertical
distribution of the condensates was not obvious (Lunine et al. 1989).
Condensates might be well mixed in the atmosphere above their condensation
level, or might coalesce into large particles, fall below their condensation
level and be removed from the atmosphere.  Of course many intermediate
cases are possible as well.  

After the discovery of what came to be known as the L and T dwarfs,
modelers initially focused on simple end cases.  Condensates were
either assumed to have either completely settled from the atmosphere 
or else were mixed uniformly throughout the observable atmosphere.
The prior approach works reasonably well for objects like
Gl 229 B (Allard et al. 1996; Marley et al. 1996; Saumon et al. 2000; 
Tsuji et al. 1996), while the latter works for very late
M and early L dwarfs like Kelu 1 (Ruiz, Leggett \& Allard 1997).  Neither 
approach, however, could adequately reproduce the colors, let alone
the spectra of
the latest Ls or the `transition' late L/early T objects, like SDSS 1254 
(Fig. 1).
\begin{figure}
 \plotone{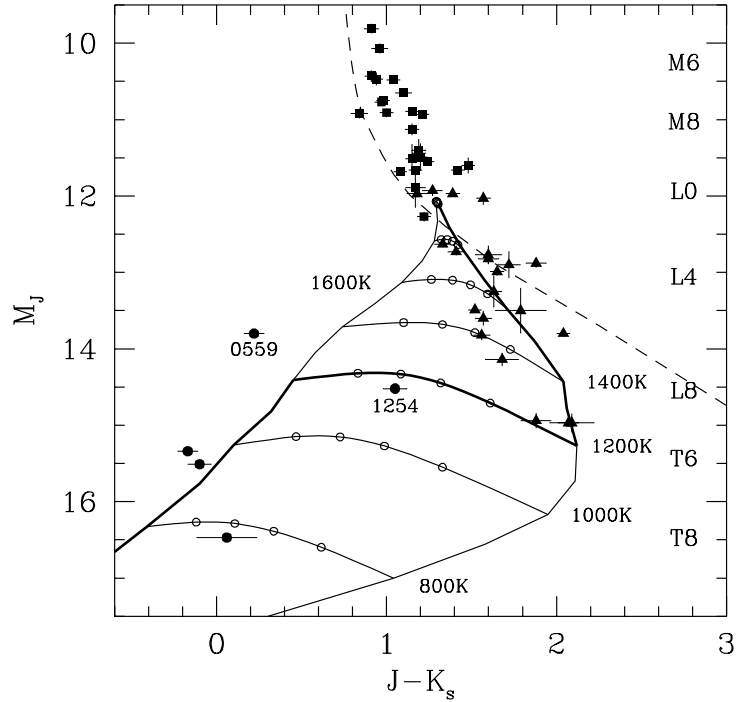}
\caption{Near-infrared color-magnitude diagram of M, L, and T dwarfs. The
absolute $J$ magnitudes and $J-K_s$ infrared colors are shown for a
sample of M (filled squares), L (filled triangles), and T 
(filled circles) dwarfs with
known parallaxes. The positions of 2MASS 0559 and 
SDSS 1254 are indicated. The
predicted colors and magnitudes for the DUSTY (dashed line),
clear (left thin line), and cloudy (right thin line; $f_{\rm sed} = 3$)
(the variable formerly known as $f_{\rm rain}$)
atmosphere models are plotted as a
function of $T_{\rm eff}$ at constant gravity, $g = 10^5\,\rm cm\, s^{-2}$
 (typical for very low mass main-sequence stars and evolved brown 
dwarfs). Connecting the cloudy and clear
 tracks are the predicted fluxes for partly cloudy models
at $T_{\rm eff} = $800, 1000, 1200, 1400, 1600, and 1800 K. 
The circles indicate the cloud
coverage fraction in steps of 20\%. The apparent evolutionary track 
of brown dwarfs based on the
 empirical data is indicated by the thickened line, which
crosses from the cloudy to clear track at $T_{\rm eff} \sim 1200\,\rm K$. 
Figure adapted from Burgasser et al. (2002).}
\end{figure}
Models (Fig. 1) in which the condensates are absent from the atmosphere
produce $J-K$ colors that are much bluer than the observed L dwarfs.  The
blue color arises from water, pressure-induced $\rm H_2$, and for lower
effective temperatures ($T_{\rm eff} < 1400\,\rm K$) 
$\rm CH_4$ absorption in $K$ band.  Models
with condensates distributed uniformly through the atmosphere match the
early Ls in which the condensate column optical depth is small, but are 
much redder than the colors of the later L dwarfs.  The
reason is that as the cloud deck falls progressively deeper in the atmosphere
the column abundance of dust gets progressively larger. Since the cloud
is forming at higher air densities the abundance of condensates to be
mixed upwards from cloud base is larger with falling effective
temperature.  Thus an outside
observer sees an ever increasing dust optical depth until the observer 
is effectively looking at a dirt-filled atmosphere.  
Like blackbodies, such objects becomes progressively redder in $J-K$ 
as they cool.
These models cannot reproduce the observed transition from red to blue in $J-K$ 
between the L and T dwarfs.

An obvious shortcoming of the first generation models is that none of them
employed cloud decks like those seen throughout the solar system.  
Condensates in planetary atmospheres are generally neither completely
settled out of the atmosphere nor distributed uniformly to the top of the
atmosphere.  Rather they tend to exist in horizontally-extended
cloud decks.  The $\rm H_2SO_4$ clouds of Venus, stratiform water 
clouds on Earth,
the ammonia cloud decks on Jupiter, and the methane clouds in Uranus
and Neptune are all examples of this.  

To demonstrate that a vertically constrained cloud deck would qualitatively
be more in line with the available data, Marley (2000) constructed a simple
model in which all clouds were 1 scale height thick.  He showed that such
a model produced less red $J-K$ colors than well mixed models and
also naturally explained the turnover in $J-K$ color from the red L dwarfs
to the blue T dwarfs.  As the cloud forms progressively lower in the
atmosphere as the object cools, the cloud disappears while the clear atmosphere above takes on the
appearance of the cloud-free models.  Tsuji (2001) also used a 
simple model with a finite-thickness cloud to make the same point.  

Given the apparent importance of correctly modeling cloud behavior,
Ackerman \& Marley (2001) developed a more rigorous cloud model 
for use in substellar
atmosphere models.  The model attempts to capture the key
`zeroth order' physics which influence the vertical condensate abundance
and size profiles in a realistic (but 1-D) atmosphere.  This model
was used by Marley et al. (2002) and Burgasser et al. (2002) to model
the atmospheres of L and T dwarfs.  Other workers have recently turned
to the work of Rossow (1978), originally developed to study
the microphysics of clouds in planetary atmospheres,
to model cloud behavior in substellar
atmospheres.  Finally Tsuji (2002) has further developed his model in 
which the cloud top temperature is an adjustable parameter.  In Section
2 we review and compare these cloud
models.  In Section 3 we summarize the results
of Burgasser et al. in applying the Ackerman \& Marley cloud model
to L and T dwarfs 
and consider the role of dynamically-induced holes in the global 
cloud coverage.

\section{A Comparison of Cloud Models}

Calculating the opacity of condensates in an atmosphere requires
estimating the distribution of particles over space and particle
size.  Estimates of condensate opacities in brown dwarf atmospheres
have all assumed a steady state, horizontally homogeneous distribution.
Beyond those common assumptions, a number of distinct approaches 
have appeared in the literature in recent years, ranging
in complexity from
(a) the simple approach of Tsuji (2002), in which the size of
cloud particles is fixed, as are the temperatures at the
bottom and top of the cloudy layer; (b) the more physically
motivated method of Ackerman \& Marley (2001), in which the
size of cloud particles and the profile of condensate mass
are coupled through an eddy mixing assumption and a parameter
describing the sedimentation efficiency; and (c) the more
detailed approach of Cooper et al. (2002), in which a number
of time microphysical and transport time constants are evaluated
in the spirit of Rossow (1978).  We briefly describe and compare
these three approaches below.

\subsection{Tsuji (2002)}

Tsuji (2002) simply assumes that all condensate particles are
10 nanometers in radius, and are vertically located between
their condensation temperature (establishing cloud
base for each condensate, assuming solar abundances) and
a prescribed temperature corresponding to cloud top for
all species.  No mention is made of the assumed vertical
distribution of condensate mass.  Likely
possibilities include the assumption of a uniform mixing ratio
or concentration.

All particles are prescribed to have a single fixed radius,
which is described as the critical radius corresponding to
the Gibbs free energy of formation of a molecular
cluster, in which the surface tension is just offset by the
supersaturation of a condensing gas.  Assumed values
of critical supersaturations and surface tensions are not given,
so the validity of a uniform critical radius for
all species is not easily evaluated.  Tsuji argues that particles larger
than this critical radius will
grow and sediment out, and therefore only particles at this
critical radius can remain within the cloud layer.  These arguments
assume that there are no vertical motions within the cloudy
air to offset sedimentation, and that the formation of the
particle clusters does not deplete the supersaturation that
gave rise to them.  It is not clear how this first argument
is to be considered consistent with other elements of the
atmospheric model, in which convective velocities are shown
to vary between 10 and 80 m/s, which are strong enough to
offset the sedimentation of particles hundreds of microns
in radius.  Furthermore, the notion that particles in
long-lived clouds are limited to their critical radius is
clearly invalid in the only atmosphere in which we have
direct cloud microphysical measurements. Water cloud 
particles in the terrestrial troposphere typically range in modal
size from 10 to 100 microns. 
In the terrestrial stratosphere sulfuric acid cloud
particle sizes are of order microns, still over one hundred times that
assumed by Tsuji.

Condensates (such as enstatite) that form at levels
cooler than the prescribed cloud top temperature are assumed to not
exist in the model atmospheres.  Tsuji suggests that such
condensates nucleate on other condensed species, and
therefore rapidly sediment out, a process akin to
seeding terrestrial cumulus clouds with large condensation
nuclei to enhance precipitation.  However, depending upon the
meteorological and background aerosol conditions, increasing
the concentration of condensation nuclei can instead
suppress precipitation in terrestrial clouds (e.g., Ackerman
et al.\ 1993).  This second possibility is ruled out by Tsuji's
model.

For the same cloud top temperatures, Tsuji finds that cloud
opacities in the atmospheres of T dwarfs are greatly reduced
from that in L dwarfs because the clouds that form
in the former (colder) atmospheres are limited (by virtue of
his cloud-top temperature cutoff) to levels below the
photosphere.  This result effectively reproduces the
result of earlier models that assume cloud layers are limited
to one scale height in depth (e.g., Marley et al., 2000).

Tsuji's choice of cloud top temperature does reproduce
the limiting value of $J-K\sim 2$ seen in the latest L dwarfs as
well as several other observed trends in the available data (Tsuji 2002).  
However since his model photometry is not combined
with an evolution model to provide radii, hence absolute magnitudes,
it is not clear if these `Unified' models are able to reproduce the
observed rapid transition in $J-K$ from the Ls to the Ts discussed in Section 3.

\subsection {Ackerman \& Marley (2001)}

The treatment of Ackerman \& Marley has been inaccurately described
(by Cooper et al., 2002) as being based on microphysical time scales.
Instead the microphysical time scale approach pioneered by Rossow (1978),
in which highly uncertain estimates are made for a number of 
microphysical processes, was bypassed by Ackerman \& Marley
in favor of a much simpler approach, in which a steady-state balance
between sedimentation (of condensate) and mixing (of condensate and
vapor) is solved at each model level.  This advective-diffusive
balance provides the profile of condensed mass in their model
atmospheres.  Condensate size distributions at each model level
are represented as a log-normal distribution that is coupled to
the advective-diffusive steady-state profile.  The width
of the distribution is a free parameter (a fixed geometric standard
deviation of 2 is assumed), while the other two unknowns in the
size distribution (total number concentration and modal radius) are 
calculated from, respectively, the steady-state condensate concentration
and the sedimentation flux.
The sedimentation flux is calculated by integrating particle
sedimentation velocity over the log-normal particle size distribution.

The Ackerman \& Marley treatment of mean particle size
can be viewed as resembling an extremely simplified variant of the
Rossow time scale
approach in that the mean particle size calculation does use particle
sedimentation speeds, and Rossow's sedimentation time scale calculation also
uses particle sedimentation speeds.  Rossow divides the atmospheric
scale height by the sedimentation speed to calculate a time scale,
but ignores eddy mixing time scales in his analysis, and instead
compares the sedimentation time scale to time scale estimates for
particle condensation, coagulation, or gravitational coalescence.
In contrast, Ackerman \& Marley effectively compare particle sedimentation
speed to the convective velocity scale to calculate mean particle
size.  

A notable extension of the Ackerman \& Marley approach beyond
preceding efforts along the same lines (e.g., Lunine et al., 1989) is
to incorporate a sedimentation scaling factor in their
computations of advective-diffusive balance and the mean
particle size.  This factor was originally called $f_{\rm rain}$ (for
``rain factor"), but rain is associated by some researchers in the
astrophysics field with the process that has been termed ``rain-out", 
in which all
condensate is removed from a saturated layer, and furthermore rain
is a term specific to condensed water.  Hence, a more 
appropriate description of the scaling factor would be ``sedimentation
factor" or $f_{\rm sed}$.  Ackerman \& Marley found that 
$f_{\rm sed} = 3$ reasonably
reproduces the observations of Jupiter's ammonia ice cloud, and
results in a condensate opacity scale height of approximately 1/4 of
a pressure scale height.  

The Ackerman \& Marley model has a small number of free
parameters, but only $f_{\rm sed}$ has been adjusted to fit observations
in practice.  The other free microphysical parameters, the
geometric standard deviation of the particle size distributions and 
the supersaturation
that persists after condensation, are fixed at 2 and 0, respectively.
The remaining free parameters relate to the difficulty of calculating
eddy diffusion coefficients from mixing length theory in stable
regions of the atmosphere.  In such regions they calculate the mixing
length by scaling the atmospheric scale height by the ratio of
the local to the adiabatic temperature lapse rate, with a minimum
scaling fixed at 0.1.  They also specify that the minimum eddy
diffusion coefficient is fixed (currently at $10^5\,\rm cm^2/s$) 
to represent residual sources of turbulence such as breaking buoyancy waves.

\subsection {Cooper et al. (2002)}

Cooper et al. (2002) and Allard et al. (this volume) draw on the 
approach of Rossow (1978) to compute cloud models.  While 
Rossow's microphysical and transport time constant approach is appealing, it
is not without its own set of stumbling blocks.
To highlight some of the assumptions inherent in this approach 
we scrutinize in this section the cloud 
model\footnote{Comments here refer to the pre-publication version
of Cooper et al. that appeared on astro-ph on May 15, 2002.} 
of Cooper et al. (2002).

For their profiles of condensate mass, Cooper et al.\  evidently
assume that all the condensate resides within one pressure
scale height above the base of each condensate cloud.  It is
not immediately obvious from their description how the condensate
mass is distributed vertically within that scale height.
They state that all the supersaturated vapor above cloud base
condenses, as assumed by Ackerman \& Marley (2001).  Recall that Ackerman \&
Marley calculate vertical profiles through the
assumption of advective-diffusive steady-state.  Cooper et al.
calculate profiles of convective velocity scale
(though unlike Ackerman \& Marley, they do not give their
formulation), which they use to calculate convective time scales
to estimate particle sizes (described below).  However, Cooper
et al. do not state whether or not (or how) they might use
the convective velocity scale to calculate their vertical 
distribution of condensate mass.  It may be the case that they
are calculating an advective-diffusive steady-state profile,
as done by Ackerman \& Marley, except that Cooper would presumably
assume all the vapor to be condensed and that 
$f_{\rm sed} = 1$, which would be somewhat consistent with
a condensate scale height matching the pressure scale height. 
However, such a profile would not be entirely consistent with their
description, which states that 100\% of the available vapor condenses
within one scale height of cloud base.  Perhaps the condensate
profile within that scale height assumes that the vapor plus
condensate is well mixed (uniform mixing ratio).  Or perhaps they
use the method of Lewis (1969) without stating so directly.
As described further in the review of other models by Ackerman
\& Marley, the approach of Lewis is easiest to understand in
terms of a parcel in an updraft, in which all the condensate
sediments at exactly the updraft speed.  Ackerman \& Marley
show that the Lewis model of Jupiter's ammonia cloud gives the
equivalent condensate profile for the Ackerman \& Marley 
model with $f_{\rm sed}\approx 5$.  Hence if
Cooper et al. are using the method of Lewis (1969) to
calculate their condensate profile, their clouds are more
physically compact than those of the baseline Ackerman and
Marley model (with $f_{\rm sed} = 3$), in contrast to the
opposite conclusion stated by Cooper et al.

To calculate the size of condensate particles, Cooper et al.
evidently assume a monodisperse distribution as done by Tsuji, rather
than distributing the condensate over a range of sizes as done
by Ackerman \& Marley.  Our only evidence of this assumption is
their omission of any mention of droplet size distributions.
Further unlike Ackerman \& Marley but like Tsuji,
Cooper et al. also assume
that the particle size is fixed with altitude, as stated in
their Section 6.3, although they show a profile of particle sizes
in their Figure 5, which is somewhat puzzling.

Cooper et al. (2002) essentially apply a hybrid approach toward
calculating the size of condensate particles.  In convective
regions they effectively calculate their particle size through
a simplified form (by virtue of their monodisperse size
distribution)
of the treatment by Ackerman \& Marley (2001), in which Cooper et
al. assume $f_{\rm sed} = 1$.  In radiative regimes Cooper et al.
use a modification of the Rossow (1978) treatment to calculate
particle size.  For this detailed treatment, Cooper et al. estimate
time scales for sedimentation, heterogeneous nucleation,
homogeneous nucleation, coagulation (collision from thermal motions), and
coalescence (collisions from differential sedimentation speeds).  
Rossow argues that the condensational time scale for cloud particles
is much longer than the nucleation time scale in planetary atmospheres
because cloud particles heterogeneously nucleate on condensation
nuclei at a rapid rate compared to the cooling rate of the cloudy 
atmosphere.  Hence, Rossow ignores the nucleation time scales in
calculating characteristic particle sizes.  In contrast, Cooper et al.
consider the process of homogeneous nucleation to be important in
the water and iron clouds of brown dwarf atmospheres, and therefore
estimate nucleation time scales.  These detailed calculations require
estimating a number of additional parameters, such as surface tension,
which requires estimating whether the cloud particles are crystalline,
glassy, or liquid.  Cooper et al. do not state their assumptions regarding
this difficult issue.  Furthermore, in his treatment of nucleation and 
condensation time scales, Rossow shows that they must be considered
in tandem to calculate a growth rate.  That is, the overall condensational
time scale is equal to inverse of the sum of the growth rate of the
particle masses plus the growth rate of the total number of particles.
However, Cooper et al. treat the nucleation rate as a growth process
on its own, a divergence from the method of Rossow that is not
obviously justifiable, nor is the departure explained.

Other important parameters that must
be estimated in their approach include the maximum supersaturation,
and the coagulation and coalescence efficiencies.  Additionally, the
shape and breadth of the size distributions must be estimated to
calculate coagulation and coalescence time scales.  Cooper et al.
provide no information regarding their assumed size distributions for
these purposes,
but presumably for their estimates of coagulation and particularly
coalescence time scales they assume that their size distributions
have some breadth, since a monodisperse size distribution would lack
dispersion of terminal sedimentation speeds, and hence coalescence
will be inoperative.

\section{The L to T Transition}

Marley et al. (2002) employed the Ackerman \& Marley (2001) cloud
model to compute atmosphere models for L and T dwarfs.  Burgasser
et al. (2002) combined these models with the evolution 
calculation of Burrows et al. (1997) to prepare a $M_J$ vs $M_J - M_K$
color magnitude diagram (Fig. 1).  Since Burrows
et al. used an earlier set of atmosphere models (that utilized
the Lunine et al. (1986) dust model) as atmospheric boundary
conditions, Figure 1 is not entirely
self-consistent.  However since model radii vary relatively little over
the magnitude range plotted, this is not an important source of
error.  In the future we will present an entirely self-consistent
evolutionary calculation. 

Figure 1 compares cloudy ($f_{\rm sed}=3$) and cloud-free models 
of L and T dwarfs with parallaxes measured by the Flagstaff USNO
group (Dahn et al. 2002; Harris, this volume).  Also shown is the 
track predicted by the
well-mixed DUSTY models of Chabrier et al. (2000).  The main conclusion
to be drawn is that neither the cloud-free nor the DUSTY models fit
the colors of the L-dwarfs.  The cloudy models better fit the data,
including the saturation in $J-K$ at around 2 seen in the latest L dwarfs.
Models that do not include particle sedimentation produce redder
colors because more condensate mass remains in the atmosphere.  Models
with more efficient sedimentation, in contrast, produce colors intermediate
between the clear and the $f_{\rm sed} = 3$ case.

Interestingly Ackerman \& Marley find that the value of $f_{\rm sed}$ 
which best fits the (poorly constrained) particle
size and vertical profile of Jupiter's ammonia cloud deck  is also 3.
More sophisticated dynamical studies of giant planet and brown dwarf
atmospheric convection may ultimately shed light on the mechanisms
underlying this result.

The challenge facing any model with a global, 1-D, cloud 
is that the $J-K$ colors of brown dwarfs swing from the red to the
blue over a very small interval in effective temperature and
$M_J$.  This is shown by the data in Figure 1 as well as other parallax
data presented at the workshop by Tinney and collaborators (this volume).
Maria Zapeterio-Osario presented colors and J magnitudes of objects
in the $\sigma$ Orionis cluster (this volume) that are also consistent with the
rapid change in color.  It is very difficult for a globally uniform cloud
to make such a sudden transition since the cloud base and, more
importantly, the optical depth of the overlying gas change slowly with
effective temperature.  The $f_{\rm sed}=3$ model in Fig. 1 typifies this
relatively slow transition.  

The rapid transition suggests that something special may be happening
at the L to T transition.  One possibility might be that the global
atmospheric dynamics change in such a way to rapidly favor more efficient
particle sedimentation (larger $f_{\rm sed}$) at the transition.
A second possibility, originally suggested by Ackerman \& Marley (2001),
is that horizontal patchiness, or holes, develop in the cloud layer
at the L to T transition.  Holes, or optically thin regions in the
global cloud deck, cover less than 10\% of Jupiter's disk.
At most thermal wavelengths optical depth unity is
reached near or above these cloud tops, so the reduced cloud opacity in 
the holes
is of little consequence.  Near $5\,\rm \mu m$, however, there is a minimum
in the combined $\rm H_2O$, $\rm CH_4$, and $\rm H_2$ gas 
opacity.  Over most of the planet
the cloud deck provides an opacity `floor' at this wavelength, but
in the cloud holes flux from deeper, warmer, and thus brighter regions
can emerge.  Essentially all of Jupiter's $5\,\rm \mu m$ radiation emerges
through these `hot spots', which were first recognized in the 1960's
(Gillett, Low \& Stein 1969).
A $5\,\rm \mu m$ image of Jupiter is shown in Figure 2.
The long path lengths through the atmosphere into the hole regions
allow relatively rare species, such as $\rm PH_3$ and
$\rm GeH_4$, to be detected in Jupiter's
atmosphere (Kunde et al. 1982).  
\begin{figure}
% \plotone{f2.eps}
\plotfiddle{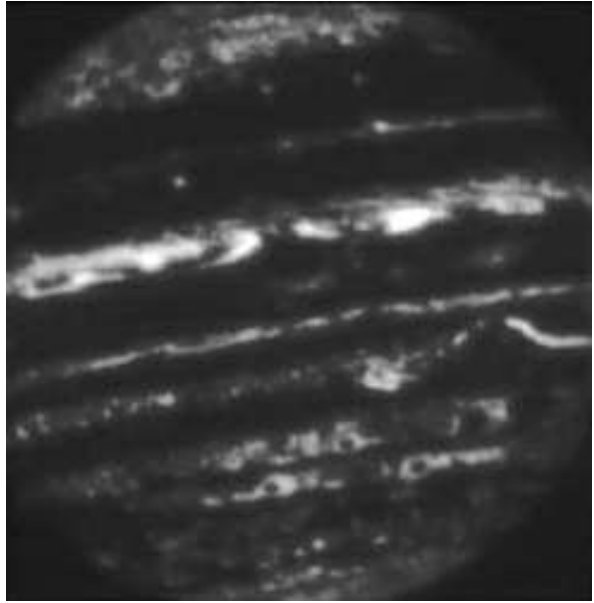}{3in}{0}{90}{90}{-100}{-9}
\caption{This image of Jupiter was taken at a wavlength centered on
$4.78\,\rm \mu m$ (narrow-band M filter) on July 26, 1995 
from the NASA Infrared Telescope Facility.  It shows thermal
emission originating from so-called ``hot spots" which are
relatively cloud-free areas in the atmosphere that allow thermal 
radiation from warmer atmospheric depths.  The Galileo
probe entered into such a region on Dec.~7, 1995 (Orton et al.
1996).  Image courtesy Glenn Orton, the IRTF, and NASA.}
\end{figure}

Burgasser et al. (2002) argue that a similar gas opacity window exists
at 1 $\micron$ in late-L/early-T dwarfs, and deeper, hotter layers
can be probed at this wavelength in breaks in the upper condensate
cloud decks.  They point to the rapid change in $J-K$ color at the
L to T transition as well as the 
observed resurgence in FeH absorption
in the earliest T dwarfs (after monotonically declining through the
L dwarfs) as supporting that clouds are patchy on brown dwarfs as well.  
In the L dwarfs gaseous
FeH abundance presumably declines as iron and silicates condense 
into clouds, leading to a simultaneous
increase in the $J-K$ color as the atmosphere becomes cloudier.  
Near the bottom of the L dwarf sequence holes begin to appear in the
clouds, allowing bright, blue (in $J-K$) flux to emerge.  This
flux pulls the integrated color over the disk blueward and can even
lead to a brightening at J band (Fig. 1).  Simultaneously FeH gas, lying below
the cloud base, again becomes detectable through the holes.  

The patchy cloud model predicts that the effective temperature range
of the earliest (T0 to T5) dwarfs is very small, the variations in
spectral properties depending more on the fractional cloud cover than
a varying effective temperature.  There are some indications that this is true
(Leggett et al., this volume), although more work is needed.  
Furthermore the patchy model
suggests that the early T dwarfs will exhibit substantial variability
in the infrared.  At this conference evidence was presented 
that the T2 dwarf SDSS 1254 is indeed apparently variable in $J$ and 
$H$ bands (Artigau et al., this volume).  

The mechanism responsible for clearings in Jupiter's cloud deck is
still not perfectly understood, although downdrafts of warm, dry
air are likely involved (Showman \& Dowling 2000).  The principle brown dwarf cloud 
decks (iron and silicate) will become subject to such vertical flows
when cloud base and the cloud tops become well seated in the atmospheric
convection zone.  In earlier type dwarfs much of the cloud opacity
lies within the statically stable radiative zone where vertical motions are
substantially smaller.  Clouds in these regions may be similar
to the stratospheric photochemical hazes of Jupiter and Titan, which 
are globally fairly uniform.

\section{Summary}
It appears that a complete description of the behavior of L and
T dwarfs will require a thorough understanding of the 
behavior of condensates in their atmospheres.  In doing so models
must simultaneously and self-consistently account for a number
of influences including particle nucleation, sedimentation, and 
vertical transport.  
Other influences, including large scale atmospheric dynamics
which may be responsible for cloud patchiness, may also be important.  
Given this daunting task, it is worth remembering that even after
half a century of dedicated effort, such key properties as the particle sizes
and vertical structure of most of the cloud decks in the solar system are
still poorly  known. The mechanisms responsible for those characteristics
which are constrained are themselves only partly understood.
Despite these challenges a coherent story for the behavior of
L and T dwarf condensates is emerging, although our understanding
is certainly still not complete.
Clouds vary in time and in space and we should perhaps
not be suprised that weather prediction is a challenging business.


\begin{references}

\reference {} Ackerman, A. S., Toon, O. B.,\& Hobbs, P. V. 1993, Science, 
262, 226.
\reference Ackerman, A.~S.~\& Marley, M.~S.\ 2001, \apj, 556, 872 

\reference{} Allard, F., Hauschildt, P.\ H.,
Baraffe, I., \& Chabrier, G.\ 1996, \apjl, 465, L123

\reference{}Burgasser, A.~J., 
Marley, M.~S., Ackerman, A.~S., Saumon, D., Lodders, K., Dahn, C.~C., 
Harris, H.~C., \& Kirkpatrick, J.~D.\ 2002, \apjl, 571, L151 

\reference{} Burrows, A.~et al.\ 
1997, \apj, 491, 856 

\reference{} Chabrier, G., Baraffe, I., Allard, F., \& Hauschildt,
P.\ 2000, \apj, 542, 464

\reference{}Cooper, C.,  Sudarsky, D., Milsom, J., Lunine, J., \& Burrows, A.
2002, astro-ph/0205192

\reference{} Dahn, C.C. et al. 2002 \aj, submitted

\reference{} Gillett, F.C., Low, F.J., \& Stein, W.A.\ 1969, \apj, 157, 925 

\reference{} Kunde, V.~et al.\ 1982, \apj, 263, 443 

\reference{} Lewis, J.~S.\ 1969, Icarus, 10, 365 

\reference  Lunine, J.~I., Hubbard, W.~B., Burrows, A., Wang, Y.-P., \& Garlow, K.\ 1989, \apj, 338, 314 

\reference{} Marley, M.\ S., Saumon, D., Guillot, T.,
Freedman, R.\ S., Hubbard, W.\ B., Burrows, A., \& Lunine,
J.\ I.\ 1996, Science, 272, 1919

\reference{}Marley, M.~S., Seager, 
S., Saumon, D., Lodders, K., Ackerman, A.~S., Freedman, R.~S., \& Fan, X.\ 
2002, \apj, 568, 335 


\reference Marley, M.\ 2000, ASP Conf.~Ser.~212: From Giant Planets to Cool Stars, 152 

\reference Orton, G.~et al.\ 1996, Science, 272, 839 

\reference{} Rossow, W.~B. 1978, Icarus, 36, 1

\reference Ruiz, M.~T., Leggett, S.~K., \& Allard, F.\ 1997, \apjl, 491, L107 

\reference{} Saumon, D., Geballe, T.\
R., Leggett, S.\ K., Marley, M.\ S., Freedman, R.\ S., Lodders, K., Fegley,
B., \& Sengupta, S.\ K.\ 2000, \apj, 541, 374

\reference{} Showman, A.~P.~\& Dowling, T.~E.\ 2000, Science, 289, 1737 

\reference Stevenson, D.~J.\ 1986, Astrophysics of Brown Dwarfs, 218 

\reference{} Tsuji, T., Ohnaka, K., Aoki, W., \& Nakajima, T.\ 1996, \aap, 308,
L29

\reference{}Tsuji, T.\ 2001, Ultracool Dwarfs: New Spectral Types L and T, 9 

\reference{}Tsuji, T.\ 2002, \apj,  in press, astro-ph/0204401



\end{references}
\end{document}